# Transverse Quantum Confinement in Metal Nanofilms: Optical Spectra

Igor Khmelinskii[1] and Vladimir I. Makarov[2]

[1]Universidade do Algarve, FCT, DQF, and CIQA, 8005-139, Faro, Portugal

[2]Department of Physics, University of Puerto Rico, Rio Piedras, P.O. Box 23343, San Juan, Puerto Rico 00931-3343, USA

*Corresponding author*: Dr. Vladimir Makarov

*Contact information*: Department of Physics
University of Puerto Rico, Rio Piedras Campus
PO Box 23343, San Juan PR 00931-3343, USA

*Phone*: 1(787)529-2010
*Fax*: 1(787)756-7717
*E-mail*: vmvimakarov@gmail.com

**Abstract**

We report optical absorption and photoluminescence spectra of Au, Fe, Co and Ni polycrystalline nanofilms in the UV-Vis-NIR range, featuring discrete bands resulting from transverse quantum confinement. The film thickness ranged from 1.1 to 15.6 nm, depending on the material. The films were deposited on fused silica substrates by sputtering/thermo-evaporation, with Fe, Co and Ni protected by a $SiO_2$ film deposited on top. The results are interpreted within the particle-in-a-box model, with the box width equal to the mass thickness of the nanofilm. The transverse-quantized energy levels and transition energies scale as the inverse square of the film thickness. The calculated values of the effective electron mass are 0.93 (Au), 0.027 (Fe), 0.21 (Co) and 0.16 (Ni), in units of $m_o$ – the mass of the free electron, being independent on the film thickness. The uncertainties in the effective mass values are ca. 2.5%, determined by the film thickness calibration. The second calculated model parameter, the quantum number $n$ of the HOMO, was thickness-independent in Au (5.00) and Fe (6.00), and increased with the film thickness in Co (7.0 to 9.0) and Ni (7.0 to 11.0). The transitions observed in absorbance all start at the level $n$ and correspond to $\Delta n = +1, +2, +3$, etc. The photoluminescence bands exhibit large Stokes shifts, shifting to higher energies with the increased excitation energy. The photoluminescence quantum yields grow linearly with the excitation energy, showing evidence of multiple exciton generation. A prototype Fe-$SnO_2$ nanofilm photovoltaic cell demonstrated at least 90% quantum yield of photoelectrons at 77K.

**Introduction**

Lately, a lot of interest was created around quantum confinement (QC) effects in different materials, with numerous publications in this area; see for example [1]. Three-dimensional, two-dimensional and one-dimensional QC has been observed in quantum dots, quantum rods, and quantum films, respectively [1, 2].

Quantum well structures exhibiting quantum confinement effects were investigated in metal nanofilms with a well-defined number of atomic monolayers deposited on single-crystal substrates, and studied mostly by angle-resolved photoemission in vacuum. These measurements were predominantly performed for the electrons with binding energies not exceeding 2-3 eV, and the phase accumulation model was used to describe the wavefunction reflections from the substrate-film and vacuum-film interfaces and calculate the energies of the quantum well states [3].

Earlier we reported indirect evidence for the one-dimensional transverse quantum confinement (TQC) in conductive and semi-conductive nanofilms, observed via exchange anticrossing spectra in nanosandwich structures [4]. Recently we reported direct spectral evidence for TQC in Si and $SnO_2$ semiconductor nanofilms [5]. Presently we report direct evidence for TQC in Au, Fe, Co and Ni nanofilms, in the form of their UV-Vis-NIR absorption and emission spectra.

**Experimental**

Fused silica substrates 25 mm in diameter and 1 mm thick (Esco Optics) were used to deposit nanocrystalline metal films. Commercial Au, Fe, Co and Ni (Sigma/Aldrich) were used to produce nanofilms on a commercial sputtering/thermo-evaporation

Benchtop Turbo deposition system (Denton Vacuum). The substrate temperature was 475ºC, unless stated otherwise. The film thickness was controlled by XRD [6], with the XPert MRD system (PANalytic) calibrated by standard nanofilms of the same materials. The estimated absolute uncertainty of film thickness was 2.5%; the relative uncertainties were much smaller, determined by the shutter opening times of the deposition system. Protective $SiO_2$ nanofilms were produced by pulsed laser deposition ($CO_2$ laser; 5 J/pulse) at 750 ºC [7].

The absorption and emission spectra were recorded on a Hitachi U-3900H UV-Visible Spectrophotometer and an Edinburgh Instruments FS5 Spectrofluorometer. The absorption spectra in the near-IR were recorded on a PF 2000 FTIR spectrometer (Perkin Elmer). The spectral peak maxima were located using PeakFit (Sigmaplot). The second-order polynomials were fitted and the fitting uncertainties estimated using the least-squares method implemented in the LINEST function (Microsoft Excel).

The photoelectric current response measurements were carried out using a high-pressure Xe lamp ($W = 1000$ W; Ariel Corporation, Model 66023), a monochromator (Thermo Jarrell Ash, Mono Spec/50), and a model 2182A nanovoltmeter (Keithley Instruments), all connected to a computer via GPIB interface, controlled by home-made software in the LabView programming environment (National Instruments).

The low-temperature measurements were carried out at 77 K using an Optistat DN-V2 optical cryostat (OXFORD Instruments). All measurements were made at 77 K, unless expressly stated otherwise.

**Results and discussion**

**Au nanofilms**

Au nanocrystalline films were deposited on fused silica substrates by sputtering. Figure 1 shows the UV-Vis absorption spectra of Au nanofilms 1.1 and 1.9 nm thick. The peak locations are listed in the Table 1, and were determined using the PeakFit software.

<Insert Fig. 1>

Table 1. Peaks in the UV-Vis-NIR spectra recorded at 77 K and the $E_1$ fitting parameter (with the fitting uncertainty) in Au nanofilms 1.1 and 1.9 nm thick. The $\Delta n$ values were assigned as explained in the text. The estimated uncertainties correspond to one standard deviation.

|                                    | 1.1 nm Au | 1.9 nm Au |       |       |       |
|------------------------------------|-----------|-----------|-------|-------|-------|
| $E_1$, cm$^{-1}$                   | 2697.1*   | 904.0±0.2 |       |       |       |
| $\Delta n$                         | 1         | 1         | 2     | 3     | 4     |
| $E_{n+\Delta n} - E_n$, cm$^{-1}$  | 29368     | 9943      | 21694 | 35252 | 50619 |

* Calculated by scaling with the inverse square of the film thickness.

The transition energies were interpreted using the particle-in-a-box model with infinite walls. The energy levels in such a system are given by the following equation, quadratic in the quantum number $n$:

$$E_n = \frac{h^2 n^2}{8 m^* L^2} \tag{1}$$

here $h$ is the Planck constant, $m^*$ the effective electron mass and $L$ the box width (nanofilm thickness). Thus for the transition energies we obtain, introducing the residual $\delta E$ to accommodate the experimental errors:

$$E_{n+\Delta n} - E_n = E_1\left(\Delta n^2 + 2n\Delta n\right) + \delta E \tag{2}$$

with

$$E_1 = \frac{h^2}{8m^*L^2} \tag{3}$$

We fitted a second-order polynomial (2) to the series of 4 peak maxima of the 1.9 nm film, obtaining a good-quality fit with $1-R^2 = 2\times10^{-10}$ and statistical uncertainties for the $E_1$ and $2nE_1$ coefficients below 0.02%. We chose the $\Delta n = 1$ value for the 9943 cm$^{-1}$ band so as to minimize the absolute value of the residual $\delta E$, which was made smaller than its uncertainty value. The fit thus produced $E_1 = 904$ cm$^{-1}$ and the starting quantum number $n = 5.00$. This quantum number corresponds to the HOMO orbital of the discrete transverse-quantized level system. Apparently the $\Delta n = \pm 1$ selection rule of the particle-in-a-box model does not apply any more; we interpret this as an indication that the transitions are occurring between adjacent atomic chains that have sufficient overlap between their wavefunctions. This latter assumption also explains why the transitions polarized in the film plane are allowed, while only the transverse-polarized transitions would be allowed in the original particle-in-a-box model. We explain the absence of the transitions from the $n$-1 *etc.* states by their strong mixing with the in-plane conduction-zone levels, leading to vanishing transition intensities (see Fig. 2a). We therefore conclude that the only observable transitions are those occurring from the level $n$ upwards, with $\Delta n = +1, +2$ etc.

<Insert Fig. 2 >

In any case, we expect that future rigorous quantum mechanical calculations will produce adequate *ab initio* values of the transition energies and transition strengths in nanofilms. Using the $L = 1.9$ nm Au film thickness, we estimate the effective electron mass $m^* = 0.93m_o$, $m_o$ being the free electron mass. The known values for thicker Au layers are $m^* = (1.00\pm0.03)m_o$ [8]. Scaling the $\Delta n = 1$ transition energy in the 1.9 nm film by the inverse square of the film thickness, we reproduce the location of the $\Delta n = 1$ band in the 1.1 nm film within 1%, corresponding to 0.5% relative uncertainty in the film thickness. Apparently, the simple model we use is describing the energy levels and transitions with high precision, as we shall equally see in Fe films.

**Fe nanofilms**

Nanocrystalline Fe films were deposited on fused silica substrates by sputtering, and protected from oxidation by $SiO_2$ nanofilms. Figure 3 shows selected absorption spectra, and Table 2 lists the spectral maxima of the Fe nanofilms in the 7.8 – 15.6 nm thickness rage, recorded at 77K.

<Insert Fig. 3>

Note that the band positions and intensities remained unchanged when the substrate was placed at 45º to the probe beam (data not shown), ruling out any eventual optical interference effects of the sample or the surface plasmon resonance phenomena, both depending on the angle of incidence.

Table 2. Measured UV-Vis absorption band maxima (cm$^{-1}$) in Fe nanofilms of different thickness. The $\Delta n$ values were assigned as explained in the text. The estimated uncertainties correspond to one standard deviation.

|  | Film thickness, nm | | | | | | |
|---|---|---|---|---|---|---|---|
|  | 7.8 | 8.3 | 9.1 | 10.3 | 11.2 | 12.2 | 15.6 |
| $E_1$, cm$^{-1}$ | 1862‡ | 1645‡ | 1368‡ | 1068* | 903.5* | 761.3±0.1 | 465.59±0.05 |
| $\Delta n$ | Absorption band maxima, cm$^{-1}$ | | | | | | |
| 1 | 24211 | 21385 | 17782 | 13883 | 11741 | 9895 | 6052 |
| 2 |  | 46047 | 38302 | 29902 | 25289 | 21313 | 13035 |
| 3 |  |  |  | 48057 | 40644 | 34254 | 20950 |
| 4 |  |  |  |  |  | 48717 | 29795 |
| 5 |  |  |  |  |  |  | 39572 |
| 6 |  |  |  |  |  |  | 50280 |

*Uncertainty estimates unavailable, as only 3 bands were recorded and (exactly) fitted by the second-order polynomial (Eq. 2).
‡Calculated by scaling with the inverse square of the film thickness.

We fitted the absorbance band maxima for the samples 12.2 and 15.6 nm thick by the second-order polynomial, Eq. 2, with $1-R^2 < 6\times10^{-10}$ and statistical uncertainties for the $E_1$ and $2nE_1$ coefficients below 0.01%. The $\Delta n$ value for the lowest-energy recorded band in each sample was chosen so as to minimize the respective $\delta E$ absolute values, which were smaller than their respective standard deviations. The fits produced the quantum number of the HOMO level $n = 6.00$ in the four thicker samples. Due to only 3 bands measured, no uncertainty estimates are available for the 10.2 and 11.2 nm films, while the same value of $n$ and $\delta E = 0$ were obtained. Using the film thickness values, we obtain $m^* = 0.027m_o$, with differences below 0.1% between the samples, demonstrating the very low relative uncertainties in the film thickness. Note that much larger values of $m^* \approx 8$ were determined in bulk Fe for thermal properties [9], whereas the values derived from the electron energy loss spectroscopic (EELS) data are 1.2 and 3.2 for the $s$- and $d$-electrons, respectively, still much higher than the present result [010]. The positions of the spectral band maxima in the three thinner samples are reproduced to better than 0.1% using the parameters obtained for the four thicker samples and the film thickness values, once more demonstrating the very low relative uncertainties of the film thickness values and an excellent correspondence between the model and the data.

Figure 4 shows the photoluminescence (PL) spectra of the 8.3 nm thick Fe nanofilm recorded at 77K in the 90º geometry, with the substrate at 45º to the excitation beam. The PL band maxima for several films of different thickness that were tested are listed in Table 3, exhibiting very large Stokes shifts, typical for quantum nanodots, see for example [11].

Table 3. Photoluminescence band maxima ($cm^{-1}$) measured at 77 K in Fe nanofilms of different thickness, in function of excitation energy, and the relative PL quantum yields for the 10.3 nm film. The samples were excited into the maximum of one of the absorption bands.

| Thickness, nm | Photoluminescence band maxima, $cm^{-1}$ (*relative quantum yields, normalized to the band integral obtained upon $\Delta n = 1$ excitation*) | | |
|---|---|---|---|
| | Exc. at $\Delta n = 1$ | Exc. at $\Delta n = 2$ | Exc. at $\Delta n = 3$ |
| 7.8 | 17541 | — | |
| 8.3 | 15430 | 18910 | |
| 9.1 | 14240 | 16930 | |
| 10.3 | 12483 (*1.0*, exc. at 13883 $cm^{-1}$) | 13983 (*2.0*, exc. at 29902 $cm^{-1}$) | 14733 (*3.7*, exc. at 48057 $cm^{-1}$) |

<Insert Fig. 4>

Note that the fact that PL is observable implies that the transverse level system is very weakly coupled to the in-plane band system, resulting in slow dissipation of the excitation energy into the continuum. This weak coupling should result from the wavefunctions of the two subsystems being essentially orthogonal to each other, due to the significantly differing spatial scales that define their respective properties. The PL spectrum depends on the excitation energy (Table 3 and Figure 4), thus the excited-state relaxation rate constants are comparable to the PL rate constants at 77K. Note also that the PL quantum yields are growing approximately linearly with the photon energy (Table 3, the 10.3 nm sample at different excitations). We interpret this result as an evidence for multiple exciton generation, also occurring in nanodots, as first reported in [12]. This process is energetically favorable in the nanofilms (Fig. 2b), as a larger exciton may be easily exchanged into several smaller ones, with some energy to spare, due to the interlevel distances increasing with the quantum number. Such exchange may also be facilitated by film surface irregularities, creating slightly differing level structures at neighboring locations. The determination of the absolute quantum yields was difficult, as the samples were thin films in an optical cryostat, whereas the usual quantum yield standards are solutions of organic dyes in an optical cell.

**Co and Ni nanofilms**

These nanofilms were deposited at different thicknesses onto fused silica substrates, and protected on top with $SiO_2$. The same measurement procedures and data treatment were used as described for the Au and Fe films. The optical spectra of some of these films are shown in Fig. 5, and the measured and calculated spectral parameters are listed in Tables 4 and 5, for Co and Ni respectively.

<Insert Fig. 5>

Table 4. Measured UV-Vis absorption band maxima and calculated spectral parameters in Co nanofilms of different thickness. Films deposited at 485ºC substrate temperature, and the estimated uncertainties correspond to one standard deviation.

|  | Film thickness, nm | | | |
|---|---|---|---|---|
|  | 7.3 | 8.1 | 9.2 | 11.3 |
| $E_1$, cm$^{-1}$ | 270.8±0.6 | 217.7±0.5 | 170.1±0.1 | 112.5±0.1 |
| $n$ | 6.98±0.02 | 7.99±0.02 | 8.00±0.01 | 9.015±0.014 |
| $E_n$, cm$^{-1}$ | 13199±78 | 13890±84 | 10870±25 | 9140±30 |
| $E_{n-1}$, cm$^{-1}$ | 9689±57 | 10630±65 | 8320±19 | 7225±24 |
| $m/m_o$* | 0.2102(5) | 0.2124(5) | 0.2107(2) | 0.2112(2) |
| $\Delta n$ | Absorption band maxima, cm$^{-1}$ | | | |
| 3 | 13790 | | | |
| 4 | 19469 | 17394 | 13601 | |
| 5 | 25683 | 22834 | 17850 | 12946 |
| 6 | 32446 | 28701 | 22438 | 16209 |
| 7 | 39746 | 35010 | 27371 | 19697 |
| 8 | 47589 | 41757 | 32643 | 23417 |
| 9 |  | 48931 | 38249 | 27352 |
| 10 |  |  | 44201 | 31521 |
| 11 |  |  | 50492 | 35911 |
| 12 |  |  |  | 40523 |
| 13 |  |  |  | 45363 |

* Uncertainties presented as units of the least significant digit in brackets.

Table 5. Measured UV-Vis absorption band maxima and calculated spectral parameters in Ni nanofilms of different thickness. The estimated uncertainties correspond to one standard deviation.

| | Film thickness, nm | | | | | | | | |
|---|---|---|---|---|---|---|---|---|---|
| | 5.3[*] | 7.9[***] | 8.7 | 9.8 | 10.7[***] | 11.4 | 12.1 | 12.9[***] | 15.1 |
| $E_1$, cm$^{-1}$ | 688.5 | 317.2±0.8 | 254.3±0.4 | 199.1±0.7 | 174.4±0.8 | 148.5±0.2 | 130.9±0.4 | 119.8±0.2 | 86.4±0.3 |
| $n$ | 7.00 | 8.02±0.02 | 8.01±0.01 | 9.03±0.04 | 9.00±0.05 | 8.96±0.02 | 10.00±0.04 | 10.02±0.02 | 11.05±0.05 |
| $E_n$, cm$^{-1}$ | 33733 | 20396±126 | 16299±62 | 16218±141 | 14126±178 | 11929±49 | 13079±111 | 12036±55 | 10319±96 |
| $E_{n-1}$, cm$^{-1}$ | 24783 | 15626±97 | 12482±48 | 12823±112 | 11162±141 | 9416±39 | 10593±90 | 9754±45 | 8536±79 |
| $m/m_o$[**] | 0.1568 | 0.1532(4) | 0.1576(2) | 0.1586(5) | 0.1519(7) | 0.1572(2) | 0.1583(5) | 0.1522(2) | 0.1573(5) |
| $\Delta n$ | Absorption band maxima, cm$^{-1}$ | | | | | | | | |
| 2 | 22028 | | | | | | | | |
| 3 | 35109 | 18108 | 14498 | | | | | | |
| 4 | 49567 | 25416 | 20350 | 17556 | 15339 | 13052 | | | |
| 5 | | 33355 | 26713 | 22943 | 20045 | 17044 | 16359 | 14998 | |
| 6 | | 41937 | 33577 | 28721 | 25109 | 21342 | 20405 | 18719 | 14228 |
| 7 | | 51144 | 40959 | 34911 | 30504 | 25933 | 24735 | 22677 | 17199 |
| 8 | | | 48844 | 41493 | 36268 | 30821 | 29312 | 26876 | 20337 |
| 9 | | | | 48467 | 42367 | 36012 | 34154 | 31311 | 23648 |
| 10 | | | | | | 41490 | 39258 | 35991 | 27109 |
| 11 | | | | | | 47268 | 44616 | 40902 | 30761 |
| 12 | | | | | | | 50252 | 46065 | 34574 |
| 13 | | | | | | | | | 38543 |
| 14 | | | | | | | | | 42710 |
| 15 | | | | | | | | | 47027 |

* Uncertainty estimates unavailable, as only three bands were recorded and exactly fitted by the second-order polynomial (Eq. 2).
** Uncertainties presented as units of the least significant digit in brackets.
*** Deposited at 485ºC substrate temperature; the remaining Ni films were deposited at 450ºC.

The spectral bands of Co and Ni (Fig. 5) have alternating intensities, with odd $\Delta n$ corresponding to the more intense bands. This is in line with the odd-$\Delta n$ transitions being allowed in the one-dimensional particle-in-a-box problem. As we already noted, we believe that the even-$\Delta n$ transitions become allowed due to surface irregularities of the film, causing a reduction in the local symmetry and allowing transitions between the adjacent atomic chains. The effective masses are larger than that presently determined in Fe nanofilms, although much smaller than the values calculated for Co and Ni based on EELS studies [10]. Note that the effective mass values in the Ni films are slightly different (0.158 vs 0.152), depending on the substrate temperature during deposition. We believe that the films deposited at different temperatures have different mechanical stress, which in turn changes lattice constants and the effective mass. Thus spectral data may be used to evaluate stress in thin films. Contrary to Fe, where the observed transitions always start from the same $n = 6.00$ level (the data on Au are insufficient for any conclusions), the $n$ in Co increases with the film thickness from 7 to 9, and in Ni increases from 7 to 11, linearly in the thickness ranges explored. The energy levels $E_{n-1}$ and $E_n$ (Tables 4 and 5) are measured from the bottom of the box (quantum well) and should be respectively located below and above the Fermi level $E_F$ (Fig. 2a). Thus, in the 11.3 nm Co film the $E_F$ is some $8 \times 10^3$ cm$^{-1}$ above the bottom of the quantum well, while in the 15.1 nm Ni film it is about $9 \times 10^3$ cm$^{-1}$ above the bottom of the quantum well (the last column in Tables 4 and 5). Using the published $E_F$ values in bulk transition metals [10], we deduce that the bottom of the quantum well is $87 \times 10^3$ cm$^{-1}$, $54 \times 10^3$ cm$^{-1}$, and $61 \times 10^3$ cm$^{-1}$ below the vacuum level in Fe, Co and Ni, respectively, for the thickest film studied. Here, we estimate that $E_F$ is $14 \times 10^3$ cm$^{-1}$ above the bottom of the quantum well in the 15.6 nm Fe film, using the values of Table 2 for $E_1$ and $n$ and the expression $E_n = n^2 E_1$. These estimates justify the apparently better polynomial fits achieved for Fe as compared to Co and Ni, as the studied excited states in Fe are always at least $20 \times 10^3$ cm$^{-1}$ below the vacuum level, and thus behave as if the potential walls were infinite.

**Nanofilm photovoltaic cell**

Inspired by the similarity in the behavior of Fe nanofilms with that of Si nanofilms [5], we investigated a prototype photovoltaic cell based on Fe and SnO$_2$, shown schematically in Fig. 6. Figure 7 shows the UV-Vis absorption spectra of the separate nanofilms, each on a fused silica substrate, and the spectrum of the two films one on top of the other.

<center><Insert Figure 6></center>
<center><Insert Figure 7></center>

Fig. 7 shows that the UV-Vis absorption spectrum of the stacked Fe and SnO$_2$ nanofilms shows the same absorption bands as the respective isolated nanofilms, demonstrating that transverse quantization in nanofilms operates independently in each of the stacked layers. Note that the SnO$_2$ films also exhibit TQC, with the respective spectra discussed in our previous publication [5]. We recorded the excitation wavelength dependence of the photocurrent at 77K, with the results shown in Figure 6, expressed as photocurrent quantum yield. Taking additionally into account the light lost by reflection off the cryostat windows, the measured quantum yield will reach 89% at the $\Delta n = 1$ Fe band maximum. The cell had the open-circuit voltage of 29.6 mV, which seems reasonable for such small film thicknesses. The cell behaved like a diode, with the direct-current resistances in two directions differing by a factor of 5. Apparently, we are dealing with a Shottky diode formed at the metal-semiconductor junction.

<Insert Figure 8>

We see that the excitation into the Fe transverse-quantized band produces a much stronger photocurrent than the excitation into the $SnO_2$ bands, while the excitation outside of these bands produces only very low photocurrents. Note also the second photocurrent band, appearing as a tail at the higher-energy end of the spectrum. Based on the film thickness value, the $\Delta n = 2$ band in this film is predicted at 52140 cm$^{-1}$, apparently with a higher quantum yield, in accordance with multiple exciton generation. Note that the UV-Vis absorption spectrum of the 7.8 nm film also indicates a tail of the $\Delta n = 2$ band at the higher-energy end of the spectrum, see Fig. 4.

**Conclusion**

We report transverse one-dimensional quantum confinement in metal nanofilms investigated by optical spectroscopy, showing that a simple particle-in-a-box model adequately describes the structure of the electronic levels quantized in the direction transverse to the film. We show that such techniques may be used to precisely measure the nanofilm thickness in the range dependent on the material, provided the material-specific effective electron mass is known. The photoluminescence quantum yields in Fe films grow with the excitation energy, indicating a possibility of multiple exciton generation. We also report photocurrent generation in a prototype Fe-$SnO_2$ photovoltaic cell at 77 K, with high quantum yields. These findings provide a new understanding of the physics of metal nanofilms, opening a new range of possibilities for the technology of solid-state devices.

**Acknowledgements**

I. K. is grateful to P. Stallinga for stimulating discussions.

**Conflicts of interest**

The authors report no conflicts of interest.

**Figure captions**

Figure 1. Absorption spectra of 1.1 nm and 1.9 nm Au nanofilms on a fused silica substrate, recorded at 77 K. The NIR part was recorded in a separate experiment.

Figure 2. a) Schematic level diagram and the experimentally observed transitions in metal nanofilms. The discrete levels that are below the Fermi level mix with the continuum, with the respective transitions not appearing in the spectrum. b) Multiple exciton generation: as the ladder steps become larger for larger $\Delta n$ values, the $\Delta n = 2$ excitation may be exchanged for two $\Delta n = 1$ excitations, with the excess energy liberated as phonons, etc.

Figure 3. Absorption spectra of Fe nanofilms 10.3, 12.2 and 15.6 nm thick (bottom-to-top, vertically shifted for visual separation) on a fused silica substrate, protected by a $SiO_2$ nanolayer and recorded at 77 K.

Figure 4. Fluorescence spectra of 8.3 nm Fe nanofilm at 77K: 1- absorption spectrum; 2- emission spectrum, excited into the $\Delta n = 1$ absorption band, 3- emission spectrum, excited into the $\Delta n = 2$ absorption band.

Figure 5. Absorption spectra of Co and Ni nanofilms on a fused silica substrate, protected by a $SiO_2$ nanolayer and recorded at 77 K. The spectra are vertically shifted for visual separation, with the film thickness indicated in nm.

Figure 6. The prototype photovoltaic cell: 1 – fused silica substrate (1 mm thick, 25 mm diameter); 2 – Au film (21.3 nm); 3 – Fe film (7.8 nm); 4 – $SnO_2$ film (3.9 nm); 5 – Au film (0.109 μm); 6 – Cu plate (0.75 mm); 7 – hollow plastic cylinder; 8 – Cu ring.

Figure 7. UV-Vis spectra (1) of the isolated 3.9 nm $SnO_2$ nanofilm, (2) of the isolated 7.8 nm Fe nanofilm, and (3) of the same layers combined one on top of the other, in all cases

on a fused silica substrate. The scale used for the separate $SnO_2$ film was expanded by a factor of 10. Note that the spectra of $SnO_2$ nanofilms also exhibit TQC, and were discussed in our previous publication [5].

Figure 8. The photocurrent quantum yield in the prototype Fe-$SnO_2$ photovoltaic cell in function of the photon energy. The larger current peak corresponds to the excitation into the $\Delta n = 1$ of the Fe nanofilm, and the smaller peaks – into the bands of the $SnO_2$ nanofilm. The higher-energy tail corresponds to the excitation into the tail of the $\Delta n = 2$ Fe band.

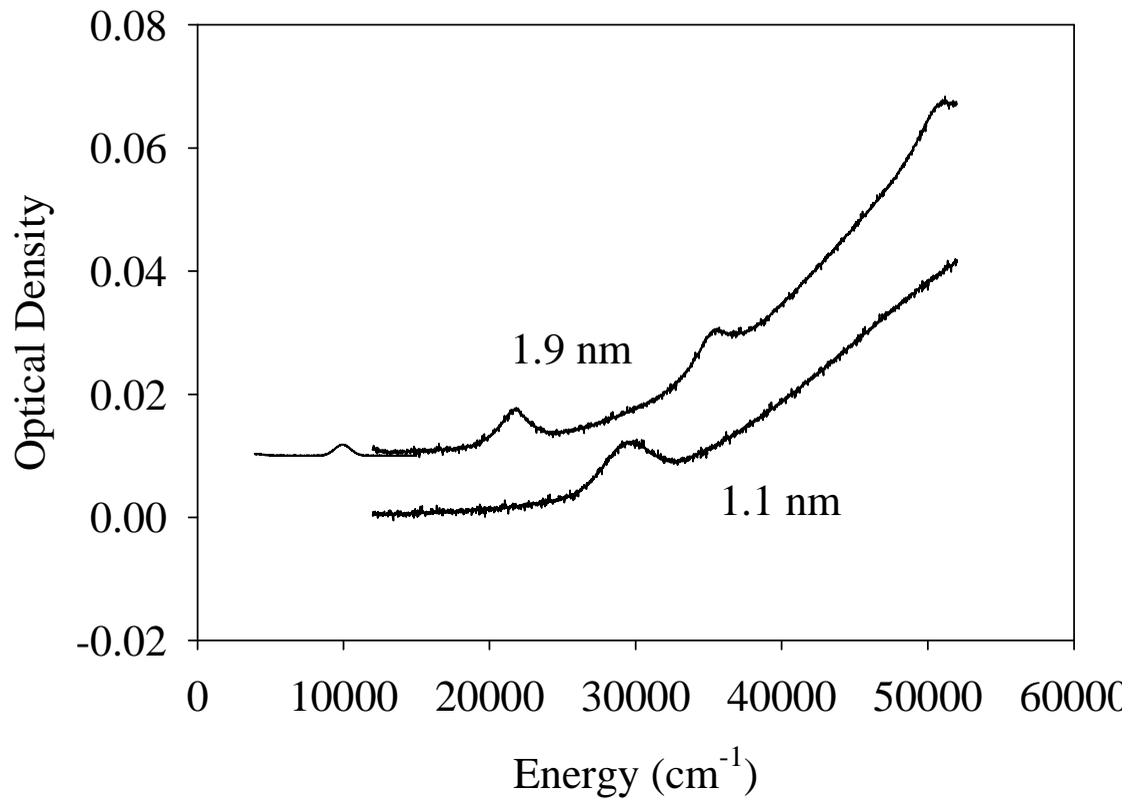

Figure 1; Khmelinskii, Makarov

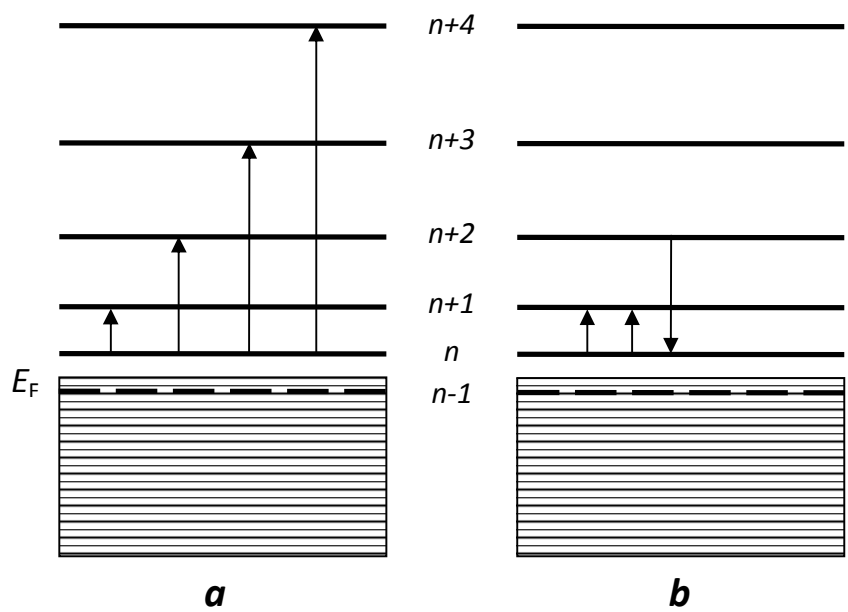

Figure 2; Khmelinskii, Makarov

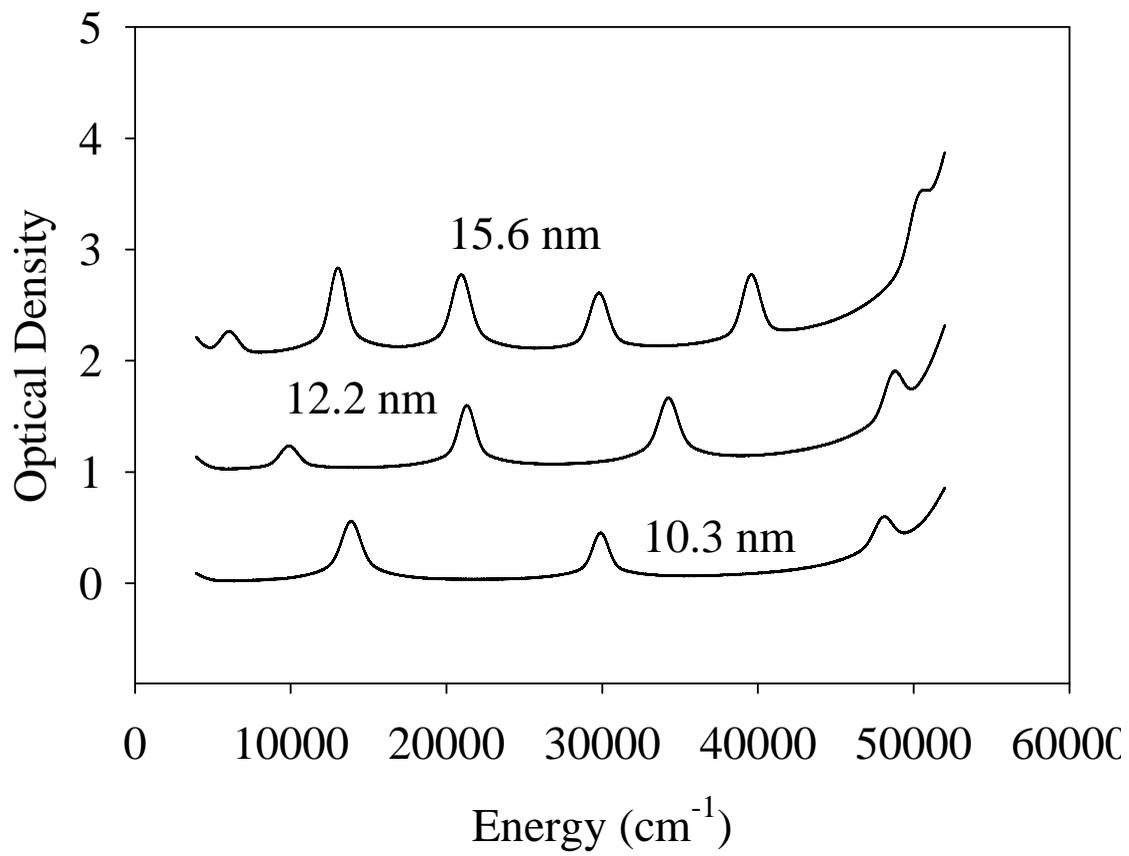

Figure 3; Khmelinskii, Makarov

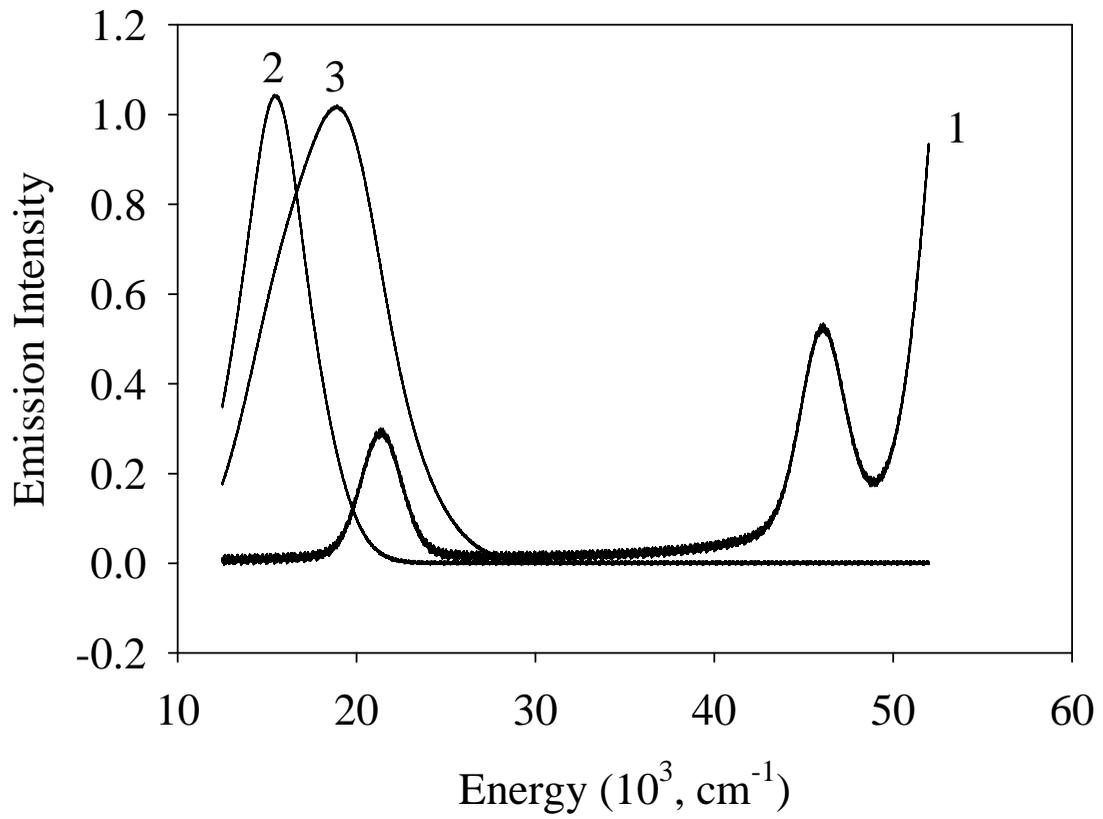

Figure 4; Khmelinskii, Makarov

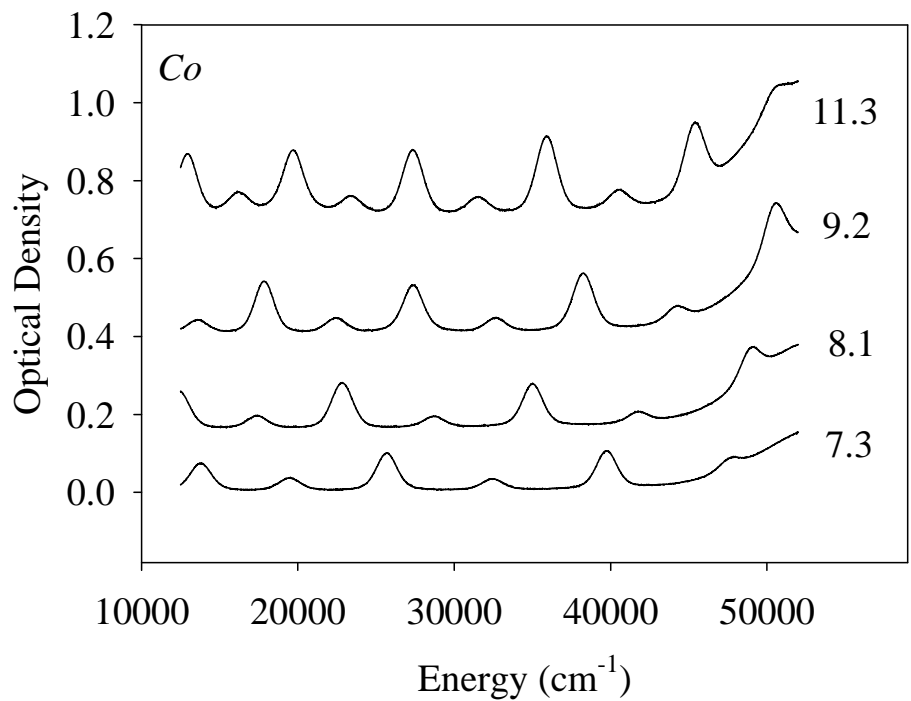

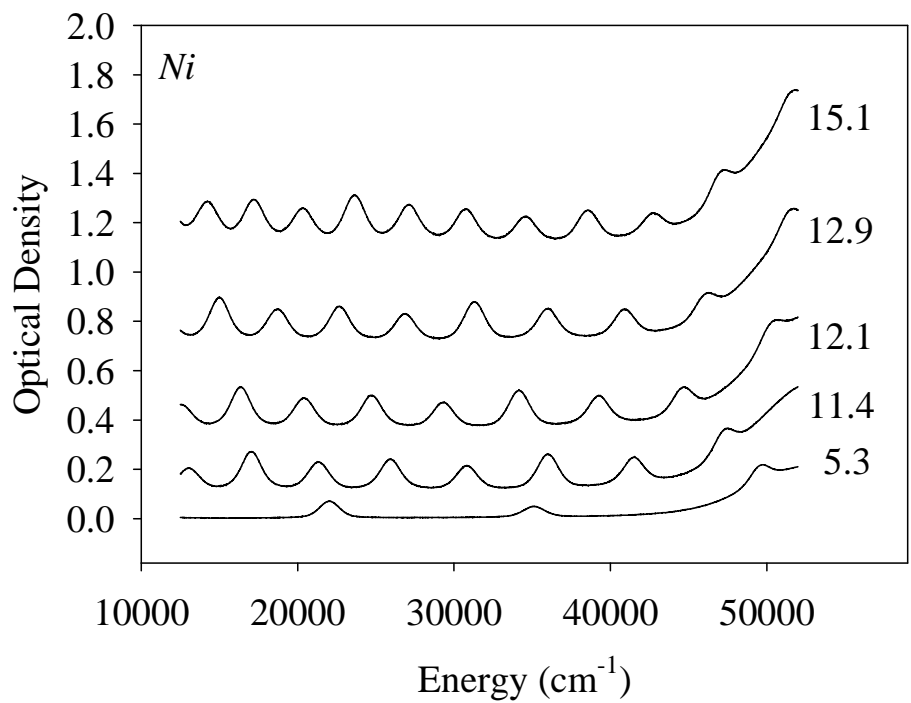

Figure 5 – Khmelinskii, Makarov

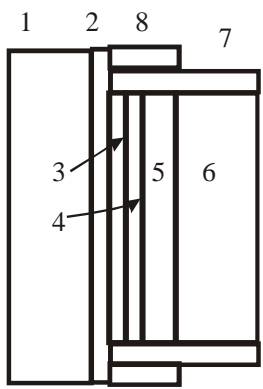

Figure 6; Khmelinskii, Makarov

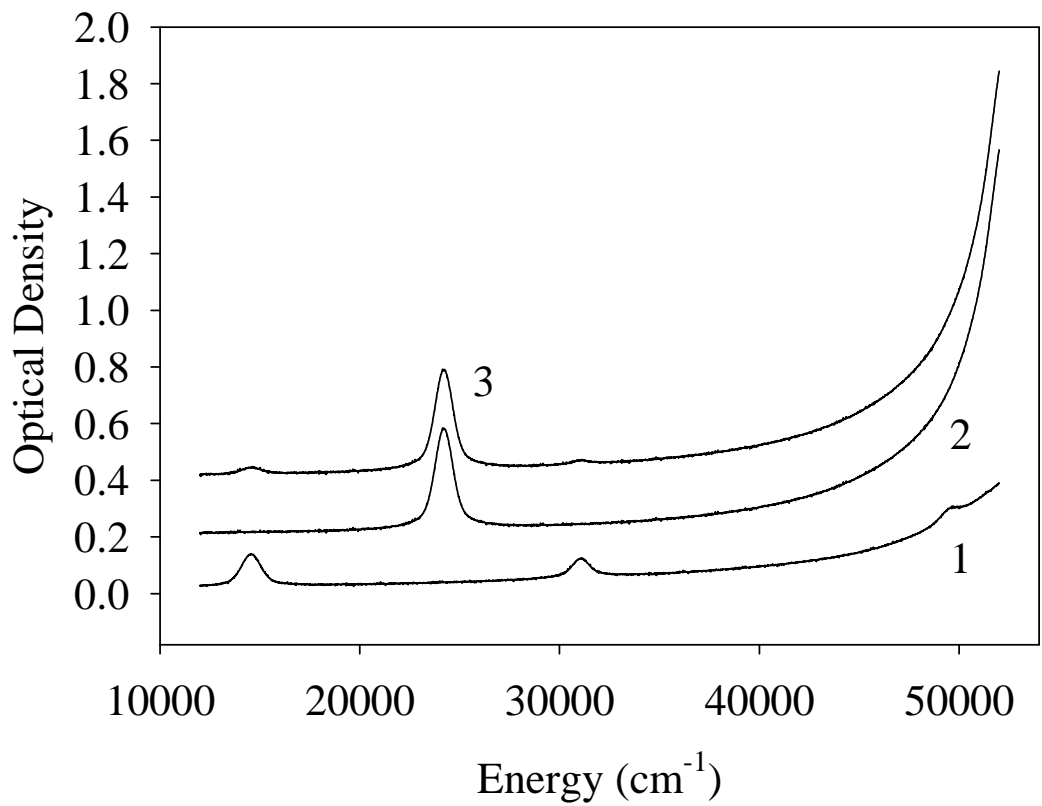

Figure 7; Khmelinskii, Makarov

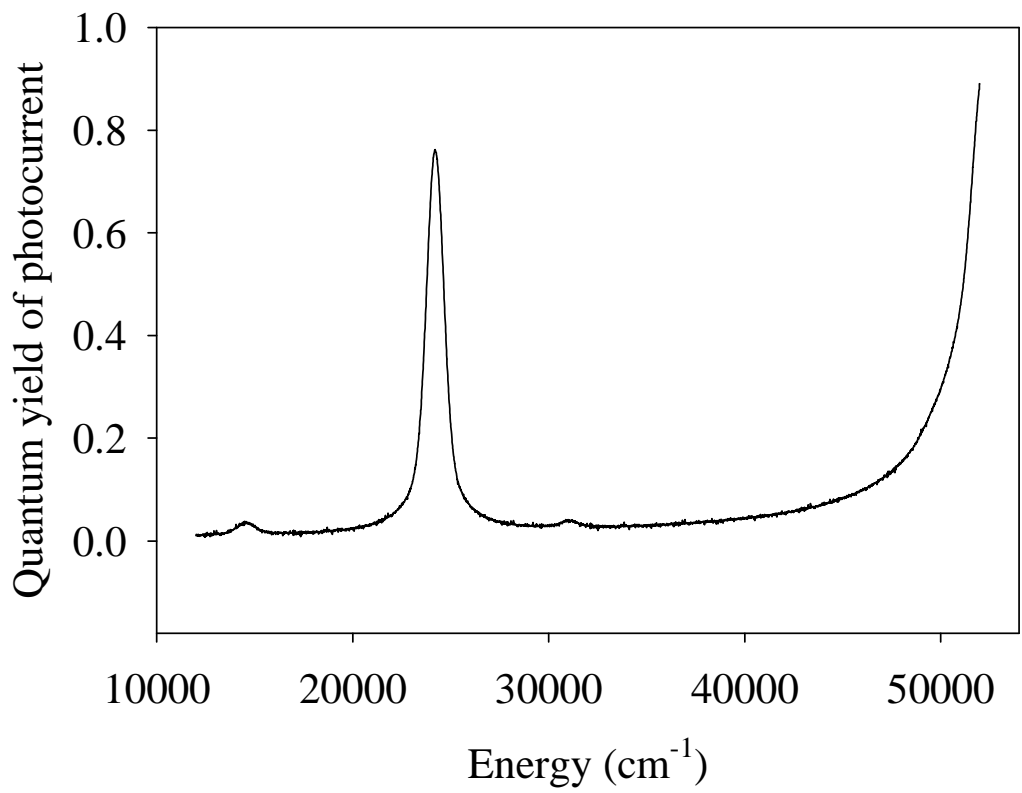

Figure 8; Khmelinskii, Makarov